\documentclass[twocolumn,10pt]{svjour3}       
\smartqed  
\usepackage{graphicx}
\usepackage{mathptmx}      
\usepackage{latexsym}
\journalname{\emph{Indian J Phys}}
\usepackage{booktabs}
\usepackage{color}
\usepackage{dcolumn}
\usepackage{bm}
\begin{document}

\title{Relativistic photoionization of H-isoelectronic series including plasma shielding effects
}

\titlerunning{Relativistic photoionization of H-isoelectronic series}        

\author{Xugen Zheng$^{1}$, Hsin-Chang Chi$^{2}$, Shin-Ted Lin$^{3}$, Gang Jiang$^{1}$, Chenkai Qiao$^{3}$ and Keh-Ning Huang$^{1,3,4,*}$ 
\\ 1 Institute of Atomic and Molecular Physics, Sichuan University, Chengdu, Sichuan, 610064, China
\\ 2 Department of Physics, National Dong Hwa University, Shoufeng, Hualien, 97401, China
\\ 3 College of Physical Science and Technology, Sichuan University, Chengdu, Sichuan, 610064, China
\\ 4 Department of Physics, National Taiwan University, Taipei, 10617, China}

\authorrunning{Xugen Zheng, Hsin-Chang Chi, Shin-Ted Lin, Gang Jiang, Chenkai Qiao and Keh-Ning Huang} 

\authorrunning{Xugen Zheng \emph{et al.}} 

\institute{*Corresponding Author, E-mail: knhuang1206@gmail.com}

\date{Received: 16 March 2018 / Accepted: 05 June 2018}

\maketitle

\begin{abstract}
With plasma shielding effects of the Debye-H\"uckel model, we investigate the relativistic photoionization processes of H, Nb$^{40+}$ and Pb$^{81+}$ plasmas in the H-isoelectronic series. The shielded nuclear potential of Yukawa-type experienced by the electron is parameterized by Debye-length $D$. To account for relativistic effects non- perturbatively, we solve the Dirac equation for the bound as well as continuum wavefunctions. Contributions from multipole fields are calculated for high incident photon energies, while the angular distribution and spin polarization parameters of photoelectrons are provided in the electric-dipole approximation. Our results of photoionization cross sections for the H plasma agree with other available theoretical calculations. The interplay between the relativistic and plasma shielding effects on the photoionization parameters is also studied.
\end{abstract}

\section{Introduction\label{sec:1}}

Spectroscopic diagnostics of laboratory and astrophysical plasmas has stimulated interests of experimental and theoretical studies in the past decades. Specifically, precise modeling for properties of plasmas demands accurate photoionization data. Debye plasmas are weakly coupled plasmas to comply with Debye-H\"uckel model with a shielding nuclear potential of Yukawa-type \cite{Debye,Margenau,Rouse,Murillo,Piel}. There is a broad category of plasmas in laboratories, astrophysical objects, and terrestrial as well as interstellar spaces, which are classified as Debye plasmas. In recent years, relativistic and non-relativistic calculations have been performed within the electric dipole and quadrupole approximations to study plasma shielding effects in the photoionization process of hydrogen-like ions submerged in Debye plasmas \cite{Weisheit,Shore,Hohne,Jung,Zhao,Qi,Lin,Chang,Xie,A1,A2}. Emphasis has been on comparative influences of plasma shielding lengths on the near-threshold photoionization process in a variety of Debye plasmas. There are also researches on the photoionization process of the H atom, hydrogen-like ions, and lithium-like ions submerged in modified Debye-H\"uckel potential or exponential-cosine-screened potential \cite{B1,B2,B3}. The relativistic and plasma screening effects on atomic structure, energy level, and atomic collisions for various kinds of screening potential have also been studied \cite{C1,C2,C3,C4,C5,C6,C7}.

In the present paper, we investigate the relativistic photoionization processes of the ground-state H atom and hydrogen-like ions Nb$^{40+}$ and Pb$^{81+}$ in Debye plasma environments for plasma diagnostics. It is noteworthy that to ionize a deeply bound electron in ions Nb$^{40+}$ and Pb$^{81+}$ required high photon energies; therefore, theoretical frameworks under the electric-dipole ($E1$) approximation will be inappropriate. With this regard, it is necessary to go beyond $E1$ approximation to include all possible multipoles that will give notable contributions. As proposed by most available theoretical investigations, we adopt the Debye-H\"uckel model to account for plasma shielding effects. The effective plasma shielded potentials are parameterized by Debye-lengths. To study the interplay between the relativistic and shielding-length effects, we have carried out calculations employing various Debye-lengths. It is pointed out that a complete analysis of photoionization processes requires the knowledge of the spin polarization as well as the angular distribution of the photoelectrons in addition to the photoionization cross section \cite{Huang1,Huang2}. In the present calculations, all significant multipole contributions for photoionization cross sections are calculated to achieve accurate total photoionization cross sections while angular distribution and total spin polarization of photoelectrons are given in the $E1$ approximation omitting the interferences arising from high-order multipoles. The angular distribution and total spin polarization parameter are provided primarily for prototypical characteristic analyses. A comprehensive study of the non-dipole interference effects on the angular distribution and total spin polarization parameters is undertaken and will appear in a following paper.

In the following sections, atomic units are employed. The theoretical method used in this paper is given in Sect. \ref{sec:2}. In Sect. \ref{sec:3}, results from present calculations including photoionization cross section, angular distribution and spin polarization parameters are presented with discussions. Conclusions are summarized in Sect. \ref{sec:4}.

\section{Theoretical Method \label{sec:2}}

\subsection{Photoionization parameters \label{sec2A}}

The basic transition matrix of photoionization process for a single-electron atomic system has the form \cite{Huang1},
\begin{equation}
T_{fi}=\frac{4\pi^{2}p_{f}E_{f}}{\omega c} \langle\Psi_{f}|\vec{\alpha}\cdot\hat{\varepsilon}e^{i\vec{k}\cdot\vec{r}}|\Psi_{i}\rangle
\end{equation}
where $\Psi_{i}$ and $\Psi_{f}$ are the initial and final state, respectively, of the single-electron system. The incident photon has the momentum $\bf{k}$ and polarization $\hat{\varepsilon}$; the outgoing photoelectron has the momentum $\bf{p}_{f}$ and energy $E_{f}$. The final state $\Psi_{f}$ of the photoelectron is normalized such that the differential cross section is given by
\begin{equation}
\frac{d\sigma_{fi}}{d\Omega}=|T_{fi}|^{2}
\end{equation}
The perturbing field can be expanded in a sum of electric and magnetic multipole terms $v_{jm}^{(\lambda)}$
\begin{equation}
\vec{\alpha}\cdot\hat{\varepsilon}e^{i\vec{k}\cdot\vec{r}}=v^{+}=\sum_{\lambda jm}v_{jm}^{(\lambda)} \label{multipole v}
\end{equation}
where the number $j$ corresponds to the $2^{j}$-pole transitions, and $\lambda$ represents the type of transition ($\lambda=E,M$ stands for the electric transition and magnetic transition respectively). Each term in (\ref{multipole v}) will induce photoionization channels with final states having the same angular momentum and parity as the perturbation. The transition amplitude from the initial state to one such final state is given as
\begin{equation}
T_{j}^{(\lambda)}= \sum_{\alpha} \langle u_{b}|v_{jm}^{(\lambda)}|u_{a}\rangle
\end{equation}
where the summation is over all possible photoionization channels allowed by the perturbation $v_{jm}^{(\lambda)}$. We use the channel index $\alpha$ to denote transition channel $a=(n_{a}\kappa_{a})\rightarrow b=(n_{b}\kappa_{b})$ associated with transitions, allowed by the perturbation $v_{jm}^{(\lambda)}$, of an electron excited from a bound orbital $u_{a}(\vec{r})$ to a continuum orbital $u_{b}(\vec{r})$. We may express the photoionization channel amplitudes in terms of reduced matrix elements, viz.,
\begin{equation}
\langle u_{b} |v_{jm}^{(\lambda)}| u_{a}\rangle
                              = \left(
                                \begin{array}{ccc}
                                  j_{b} & m & m_{a} \\
                                  m_{b} & j & j_{a}
                                \end{array}
                                \right)
                                D_{\alpha}(\lambda j)  \label{reduced matrix element}
\end{equation}
where $j_{a}$ and $m_{a}$ denote angular-momentum quantum and the magnetic quantum number, respectively, of an orbital $u_{a}(\vec{r})$. We refer the interested readers to \cite{Huang1} for furthermore descriptions of the reduced matrix elements $D_{\alpha}(\lambda j)$.

The total photoionization cross section for an electron in state $(n\kappa)$ is given by \cite{Huang1}:
\begin{equation}
\sigma_{n\kappa}=\frac{4\pi^{4}c}{\omega(2j_{0}+1)} \bar{\sigma}_{n\kappa}
\end{equation}
where
\begin{eqnarray}
\bar{\sigma}_{n\kappa} & = & \sum_{\lambda j \alpha}D^{2}_{\alpha}(\lambda j) \nonumber \\
                       & = & \sum_{j\alpha}[D^{2}_{\alpha}(Ej)+D^{2}_{\alpha}(Mj)]  \label{reduced cross section}
\end{eqnarray}
Here $D_{\alpha}(Ej)$ and $D_{\alpha}(Mj)$ are the photoionization reduced matrix elements corresponding to channels $\alpha$ arising from electric and magnetic $2^{j}$-pole excitations, respectively. In the electric-dipole approximation, it is conventional to abbreviate $D_{\alpha}(E1)$ using the shorthand notation $D_{j_{\alpha}}\equiv D_{\alpha}(E1)$, where $j_{\alpha}$ is the total angular momentum of the photoelectron in channel $\alpha$.

\begin{figure}
\includegraphics[width=0.52\textwidth]{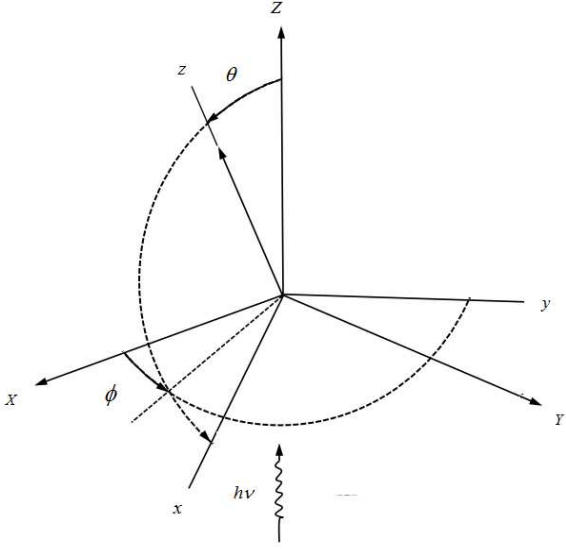}
\caption{\label{coordinate} Coordinate system $XYZ$ and $xyz$.}
\end{figure}

The angular distribution and spin polarization of photoelectrons have been derived for an arbitrarily polarized incident photon including all multipole transitions \cite{Huang1}. As a simple example, under the electric-dipole approximation for circular polarized incident photon, the differential cross section and spin polarization of photoelectrons are given by \cite{Huang1}
\begin{eqnarray}
\frac{d\sigma_{n\kappa}}{d\Omega} &=& \frac{\sigma_{n\kappa}}{4\pi}
                                      [1-\frac{1}{2}\beta_{n\kappa}P_{2}(\cos{\theta})] \label{differential1}
\\
P_{x}(\theta,\phi) &=& \frac{\pm\xi_{n\kappa}\sin{\theta}}{1-\frac{1}{2}\beta_{n\kappa}P_{2}(\cos{\theta})} \label{polarization1}
\\
P_{y}(\theta,\phi) &=& \frac{\eta_{n\kappa}\sin{\theta}\cos{\theta}}{1-\frac{1}{2}\beta_{n\kappa}P_{2}(\cos{\theta})} \label{polarization2}
\\
P_{z}(\theta,\phi) &=& \frac{\pm\zeta_{n\kappa}\cos{\theta}}{1-\frac{1}{2}\beta_{n\kappa}P_{2}(\cos{\theta})} \label{polarization3}
\end{eqnarray}
where $n$ and $\kappa$ are, respectively, the principle and angular quantum numbers, while the $\pm$ signs are for photon with positive or negative helicity and $\theta$ denotes the angle between the momentum $\bf{p}$ of the ejected electron and the momentum $\bf{k}$ of the incident photon. The coordinate systems adopted for observations are prescribed below. We define a fixed coordinate system $XYZ$ such that the $Z$ axis is in the direction of the photon flux, and $X$ axis can be chosen in any convenient direction perpendicular to $Z$ axis. A rotated coordinate system $xyz$ is determined from the fixed coordinate system $XYZ$ by rotations with Euler angle $(\phi,\theta,0)$. The rotated coordinate system $xyz$ is chosen such that the $z$ axis, making the angle $\theta$ withe the $Z$ axis, is the direction of the outgoing photoelectron and the $y$ axis is normal to both the $Z$ and $z$ axes. The spin polarization of the photoelectron is defined with respect to the rotated coordinate system $xyz$. The relative orientation of these two coordinate systems is shown in Fig. \ref{coordinate}.

The five parameters $\sigma_{n\kappa}$, $\beta_{n\kappa}$, $\xi_{n\kappa}$, $\eta_{n\kappa}$ and $\zeta_{n\kappa}$ are inherent in the dynamical properties associated with the photoionization process. In (\ref{differential1}), $\sigma_{n\kappa}$ is the total photoionization cross section integrated over all photoelectron angles, while $\beta_{n\kappa}$ is the angular asymmetry parameter of the differential photoionization cross section. From (\ref{polarization1}) to (\ref{polarization2}), $\xi_{n\kappa}$, $\eta_{n\kappa}$ and $\zeta_{n\kappa}$ are, in turns, the spin-polarization parameters related to the spin-polarizations of photoelectrons in the $x$, $y$ and $z$ directions, respectively. As we can seen from (\ref{differential1})-(\ref{polarization3}), the angular information about the differential cross section and the spin polarization of photoelectron is incorporated into the dynamical parameter $\beta_{n\kappa}$, $\xi_{n\kappa}$, $\eta_{n\kappa}$ and $\zeta_{n\kappa}$. The total spin polarization of photoelectrons are found to be $P_{X}=P_{Y}=0$ and $P_{Z}=\delta_{n\kappa}S_{3}$, where $\delta_{n\kappa}$ is the total spin polarization parameter defined by
\begin{equation}
\delta_{n\kappa}=\frac{1}{3}(\zeta_{n\kappa}-2\xi_{n\kappa})
\end{equation}
It is noticed that $P_{Z}$ depends linearly on the Stokes parameter $S_{3}$.

\begin{figure}
\includegraphics[width=0.52\textwidth]{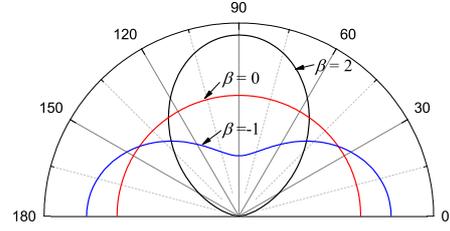}
\caption{\label{Angular Dirstribution1} Polar diagram showing $d\sigma_{n\kappa}/d\Omega$ as function of emission angle $\theta$ of photoelectrons. The radius at a specific polar angle $\theta$ indicates the magnitude of $d\sigma_{n\kappa}/d\Omega$ at that angle. The red, black and blue curves correspond to $\beta_{n\kappa}=0,2$ and -1, respectively.}
\end{figure}

The angular distribution of photoelectrons is characterized by the asymmetry parameter $\beta_{n\kappa}$. For illustration purposes, we present a polar diagram of $d\sigma_{n\kappa}/d\Omega$ as functions of emission angles $\theta$ of photoelectrons via photoionization of $s$ subshell electrons in Fig. \ref{Angular Dirstribution1} within $E1$ approximation. The radii at various angles $\theta$ represent the magnitudes of $d\sigma_{n\kappa}/d\Omega$. Since $-1\leq \beta_{n\kappa} \leq2$ owing to the requirement that $d\sigma_{n\kappa}/d\Omega$ can not be smaller than 0, here we have chosen $\beta_{n\kappa}= -1, 0$ and $2$ as representative examples. As Fig. \ref{Angular Dirstribution1} shows, the photoelectron distribution is uniform at any angle $\theta$ when $\beta_{n\kappa}=0$.  Moreover, when $\beta_{n\kappa}>0$, photoelectrons incline to appear more likely near the angles $\theta=90^{o}$. However, when $\beta_{n\kappa}<0$, photoelectrons tend to emerge more probably around the angles $\theta=0^{o}$ and $180^{o}$.

In the electric-dipole approximation, for unpolarized single-electron targets in the $^{2}S_{1/2}$ ground states, the allowed $jj$-coupling photoionization channels of the $1s$ electron are summarized below.
\begin{eqnarray}
\textrm{Channel}\ 1: && 1s \rightarrow \varepsilon p_{1/2} \nonumber
\\
\textrm{Channel}\ 2: && 1s \rightarrow \varepsilon p_{3/2} \nonumber
\end{eqnarray}
where $\varepsilon$ represents the photoelectron energy. In such cases, there are only two electric-dipole amplitudes and one relative phase; hence only three independent dynamical parameters are possible. Furthermore, it is thus legitimate to select the 3 dynamical parameters $\sigma_{1s}$, $\beta_{1s}$ and $\delta_{1s}$ to be independent. For brevity, we use the notations $D_{1/2}$ and $D_{3/2}$ to denote the reduced photoionization amplitudes corresponding to channels 1 and 2, respectively. The explicit expressions of these parameters in terms of $D_{1/2}$ and $D_{3/2}$ can be expressed as \cite{Huang1,Fano}:
\begin{eqnarray}
\sigma_{1s} & = & \frac{2\pi^{4}c}{\omega} (|D_{1/2}|^{2}+|D_{3/2}|^{2}) \label{sigma expression}
\\
\beta_{1s} & = & \frac{|D_{3/2}|^{2}+\sqrt{2}(D_{1/2}D_{3/2}^{*}+D_{3/2}D_{1/2}^{*})}
                      {|D_{1/2}|^{2}+|D_{3/2}|^{2}}
\\
\delta_{1s} & = & \frac{5|D_{3/2}|^{2}-2|D_{1/2}|^{2}-2\sqrt{2}(D_{1/2}D_{3/2}^{*}+D_{3/2}D_{1/2}^{*})}
                       {6(|D_{1/2}|^{2}+|D_{3/2}|^{2})} \nonumber
\\ \label{delta expression}
\end{eqnarray}
The above three independent parameters suffice to describe the photoionization process completely in the E1 approximation. It is worth noting that, in the non-relativistic limit, the angular asymmetry and spin-polarization parameters $\beta_{1s}$ and $\delta_{1s}$ will attain constant values 2 and 0, respectively. In cases of high incident photon energies while multipole effects are significant, it is also worthwhile to point out that the interferences among multipole transition amplitudes arising from photoionization channels induced by different multipoles must be carefully accounted for to achieve accurate $\beta_{1s}$ and $\delta_{1s}$.

\subsection{Debye-H\"uckel model \label{sec2B}}

\begin{figure}
\includegraphics[width=0.55\textwidth]{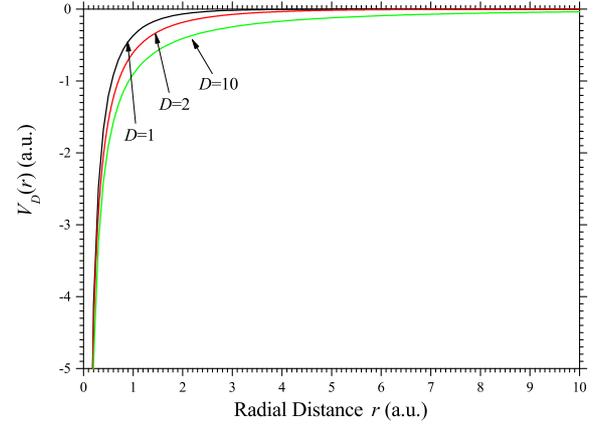}
\caption{\label{Debye potential} Debye-H\"uckel potentials with different shielding Debye lengths in neutral H atoms.}
\end{figure}

\begin{figure}
\includegraphics[width=0.55\textwidth]{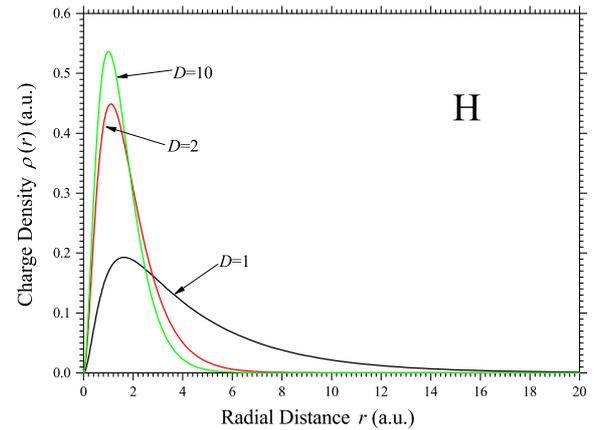}
\caption{\label{Charge distribution} The charge distribution of $1s$ electron in H atom embedded in Debye plasmas with Debye length $D = 1$, $D = 2$ and $D = 10$.}
\end{figure}

A wide group of laboratory and astrophysical plasmas are Debye plasmas. For Debye plasmas, the electron potential of a single-electron atomic system is given by
\begin{equation}
V_{D}(r)= -\frac{Z}{r}e^{-r/D} \label{potential}
\end{equation}
within the Debye-H\"uckel model\cite{Debye,Margenau,Rouse,Murillo,Piel}. In (\ref{potential}), $Z$ is the nuclear charge and $D$ donates the Debye length, respectively. The Debye length $D$ is proportional to the square root of the electron temperature divided by electron density. To visualize the plasma shielding on the nuclear potentials using hydrogen as an example, we depict the plasma shielded potentials with different Debye lengths $D = 1$, $D = 2$ and $D = 10$ in Fig. \ref{Debye potential}.  As Fig. \ref{Debye potential} shows, short Debye length manifests stronger plasma shielding on the nuclear charge in Debye plasmas. A Debye plasma with infinite Debye length is indeed equivalent to a free atom, i.e. an unshielded atomic system with Coulomb nuclear potential. Furthermore, Fig. \ref{Debye potential} indicates that discrete states are not supported for atomic systems imbedded in Debye plasmas with minuscule Debye lengths since the Coulomb nuclear potential is highly shielded off. In Fig. \ref{Charge distribution}, we plot the corresponding charge distributions associated with the individual shielded potentials given in Fig. \ref{Debye potential}. It consistently shows that the loosely bounded atomic electron is attracted toward the nuclear because of the strong shielding of the nuclear potential by the plasma surroundings.

\subsection{Wave functions \label{sec2C}}

In this subsection, we give a theoretical account for obtaining the bound state and continuum state wavefunctions. Our approach is based on the relativistic single-electron Hamiltonian incorporated with Debye-H\"uckel model
\begin{equation}
H=c\vec{\alpha}\cdot\vec{p}+(\beta-1)c^{2}+V_{D}(r)
\end{equation}
where $V_{D}(r)$ is Debye-H\"uckel potential given in Eq. (\ref{potential}). The orbital wavefunctions $u_{a}(\vec{r})$ are assumed to be in the central-field form
\begin{equation}
u_{a}(\vec{r})=\frac{1}{r}
               \left( \begin{array}{c}
                 G_{n_{a}\kappa_{a}}(r)\Omega_{\kappa_{a} m_{a}}(\theta,\phi) \\
                 iF_{n_{a}\kappa_{a}}(r)\Omega_{-\kappa_{a} m_{a}}(\theta,\phi)
               \end{array} \right)
\end{equation}
where $a$ denotes the quantum numbers $a=(n_{a}\kappa_{a})$, and the angular functions $\Omega_{\kappa m}$ are normalized spherical spinors. The normalized spherical spinors are defined as
\begin{equation}
\Omega_{\kappa m}=\Omega_{jlm}=\sum_{M\mu} \langle lM\frac{1}{2}\mu|jm\rangle Y_{lM}(\hat{r})\chi_{\mu}
\end{equation}
where $Y_{lM}$ is the spherical harmonics, and $\chi_{\mu}$ the spinor with $s= 1/2$ and $s_{z}=\mu$.

We introduce the two-component radial orbitals
\begin{equation}
u_{a}\equiv{u_{a}}(r)=\left( \begin{array}{c}
                                    G_{n_{a}\kappa_{a}} \\
                                    F_{n_{a}\kappa_{a}}
                                  \end{array} \right)
\end{equation}
and define the radial Hamiltonian operator as
\begin{eqnarray}
h_{a}\equiv h_{a}(r)= \left(
                      \begin{array}{cc}
                        V_{D}(r) & -c\bigg(\frac{d}{dr}-\frac{\kappa_{a}}{r}\bigg) \\
                        c\bigg(\frac{d}{dr}+\frac{\kappa_{a}}{r}\bigg) & V_{D}(r)-2c^{2}
                      \end{array}
                      \right)
\end{eqnarray}
where c is the speed of light. Subsequently, the radial orbital equation for orbital $u_{a}$ is given by \cite{Huang2,Matese}
\begin{equation}
(h_{a}-\varepsilon_{a})u_{a}=0 \label{orbital equation}
\end{equation}

(i) For bound state orbital with $\varepsilon_{a}<0$ , we impose the following boundary conditions for $u_{a}$:
\begin{eqnarray}
G_{a}(r=0) & = & 0 \\
F_{a}(r=0) & = & 0 \\
G_{a}(r\to\infty) & = & 0 \\
F_{a}(r\to\infty) & = & 0
\end{eqnarray}
The bound state orbitals are normalized to 1.

(ii) For continuum state orbital with $\varepsilon_{a}>0$, orbitals $u_{a}$ are subject to the following boundary conditions:
\begin{eqnarray}
G_{a}(r=0) & = & 0 \\
F_{a}(r=0) & = & 0 \\
u_{a}(r\to\infty) & {\longrightarrow} & cos\delta_{a}f_{a}
                                        +sin\delta_{a}g_{a} \label{Coulomb shift} \\
f_{a}(r\to\infty) & {\longrightarrow} & \frac{1}{c}
                                             \left( \begin{array}{c}
                                               \sqrt{ \frac{\varepsilon_{a}+2c^{2}}{\pi p_{a}} }cosX_{a} \\
                                               -\sqrt{ \frac{\varepsilon_{a}}{\pi p_{a}} }sinX_{a}
                                             \end{array} \right) \\
g_{a}(r\to\infty) & {\longrightarrow} & -\frac{1}{c}
                                             \left( \begin{array}{c}
                                               \sqrt{ \frac{\varepsilon_{a}+2c^{2}}{\pi p_{a}} }sinX_{a} \\
                                               \sqrt{ \frac{\varepsilon_{a}}{\pi p_{a}} }cosX_{a}
                                             \end{array} \right)
\end{eqnarray}
\begin{eqnarray}
X_{a} & = & p_{a}r+\frac{\mu(2p_{a}r)}{n}-\frac{(l+1)\pi}{2}+\lambda_{a} \label{non-Coulomb shift} \\
\mu & = & \frac{Z(\varepsilon_{a}+c^{2})}{cp_{a}}
\end{eqnarray}
The parameters $\delta_{a}$ and $\lambda_{a}$ in (\ref{Coulomb shift}) and (\ref{non-Coulomb shift}) correspond to the Coulomb and non-Coulomb phase shifts, individually. The continuum orbitals are normalized on the energy scale.

With the bound and continuum orbitals determined separately, the multipole photoionization amplitudes are obtained in terms of the multipole reduced matrix elements $D_{\alpha}(Ej)$ and $D_{\alpha}(Mj)$ introduced in (\ref{reduced cross section}). Explicit expressions of $D_{\alpha}(Ej)$ and $D_{\alpha}(Mj)$ suitable for numerical evaluations are presented in Appendix C of the first article in \cite{Huang1}.

\section{Results and Discussions\label{sec:3}}

In the present study, we carry out calculations beyond the $E1$ approximation to include all multipoles giving significant contributions to the total photoionization cross sections. The omitted contributions from remaining higher multipoles are estimated to be smaller than one part per ten thousand compared to the converged cross sections. In the meanwhile, the angular distribution parameter $\beta_{1s}$ and spin-polarization parameter $\delta_{1s}$ are calculated in the $E1$ approximation with interferences from all higher multipoles truncated. We present results for the photoionization cross section, as well as angular distribution and spin polarization of photoelectrons from present calculations in the following Sects. \ref{sec3A} and \ref{sec3B}, respectively. It is remarked that we estimate the relative numerical uncertainty to be at the order $10^{-6}$ in the present calculations employing a double-precision numerical scheme. For this reason, the resulted presented from our calculations are given with five significant digits.

\begin{figure}
\includegraphics[width=0.55\textwidth]{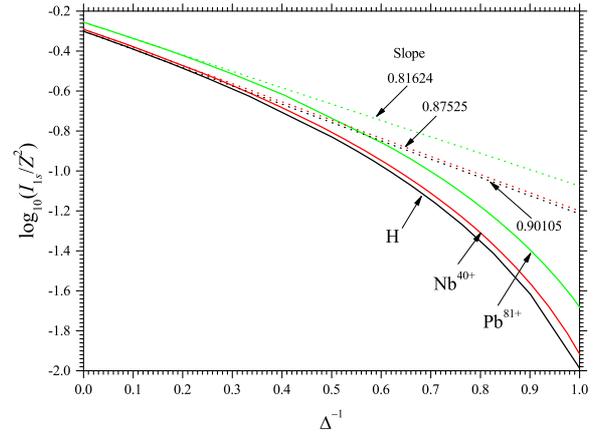}
\caption{\label{binding energy picture} Graph plotting logarithm of scaled binding energy $I_{1s}/Z^{2}$ against the inverse of Debye scaled shielding length $\Delta^{-1}$ for H atom together with H-like ions Nb$^{40+}$ and Pb$^{81+}$ ions. Solid lines are obtained from solving Dirac equations with plasma shielding effects included. Dot lines are best fitted linear functions using $\Delta^{-1}\leq0.15$ for individual solid lines. The slopes of corresponding dot lines are also given.}
\end{figure}

To demonstrate the influence of plasma shielding effects on the binding energy of the $1s$ electron, we give, in Table \ref{table binding energy}, the dependence of binding energy $I_{1s}$ on several scaled shielding lengths for H, Nb$^{40+}$ and Pb$^{81+}$. The binding energy exhibits an expected feature: as the shielding lengths being shortened, the binding energy will be diminished as well, due to the enhanced shielding off the nuclear charge by the plasma environment. In particular, $\Delta\rightarrow0$ corresponds to full shielding off the nuclear charge, the $1s$ electron becomes a free electron in consequence. In contrast, $\Delta=\infty$ corresponds to zero shielding off the nuclear charge in coincidence with a pure Coulomb instance. Since we employ a relativistic framework applying Dirac equation, the relativistic effects are taken into account from the outset. For unrevealing the effects interplayed by the shielding and relativity, we depict, in Fig. \ref{binding energy picture}, the logarithms of the scaled binding energy with respect to the inverse of the scaled shielding lengths for H, Nb$^{40+}$ and Pb$^{81+}$. Here the scaled binding energy is defined as $I_{1s}/Z^{2}$. It is evident that the logarithms of scaled binding energy depend almost linearly on the inverse of the scaled shielding length near the zero-shielding end, especially for $\Delta^{-1}<0.15$. In the linear region, we may ascribe the characteristics of binding energy to be predominantly affected by relativistic effects, showing a $Z^{2}$ dependence of $I_{1s}$ as in the Coulombic case. For $\Delta^{-1}>0.15$, as plasma shielding effects come into play, the scaled binding energy deviates from a linear relation with $\Delta^{-1}$. The shielding effects in conjunction with the relativistic effects seem to enlarge the relative difference between the binding energies of a H-like ion and neutral H atom at a certain $\Delta$, a self-explanatory evidence which we may judge from the widened separation between the Nb$^{40+}$ and Pb$^{81+}$ curves in Fig. \ref{binding energy picture}.

\begin{table}
\caption{Binding energies in a.u. for ground-state H atom and H-like ions Nb$^{40+}$ and Pb$^{81+}$ with various scaled shielding lengths $\Delta$. }
\label{table binding energy}
\setlength{\tabcolsep}{1.1mm}{
\begin{tabular}{lcccccccccc}
\hline\noalign{\smallskip}
$\Delta$ & & \multicolumn{3}{c}{$I_{1s}$} & & \multicolumn{3}{c}{$\bar{I}_{1s}=I_{1s}/Z^{2}$}
\\
\noalign{\smallskip}\hline\noalign{\smallskip}
         & & H       & Nb$^{40+}$ & Pb$^{81+}$ & & H       & Nb$^{40+}$  & Pb$^{81+}$
\\
\noalign{\smallskip}\hline\noalign{\smallskip}
1.0      & & 0.0103  & 20.5701    & 140.3718   & & 0.0103  & 0.0122      & 0.0209
\\
1.1      & & 0.0228  & 43.3350    & 257.6358   & & 0.0228  & 0.0258      & 0.0383
\\
1.2      & & 0.0372  & 69.0819    & 383.2744   & & 0.0372  & 0.0411      & 0.0570
\\
1.4      & & 0.0675  & 122.6082   & 633.5476   & & 0.0675  & 0.0729      & 0.0942
\\
1.6      & & 0.0969  & 173.8025   & 865.1625   & & 0.0969  & 0.1034      & 0.1287
\\
2.0      & & 0.1481  & 262.5175   & 1255.8698  & & 0.1481  & 0.1562      & 0.1868
\\
3.0      & & 0.2368  & 414.6706   & 1906.2227  & & 0.2368  & 0.2467      & 0.2835
\\
5.0      & & 0.3268  & 567.7840   & 2544.0468  & & 0.3268  & 0.3378      & 0.3784
\\
10.0     & & 0.4071  & 703.5982   & 3099.6242  & & 0.4071  & 0.4186      & 0.4610
\\
50.0     & & 0.4803  & 827.0424   & 3597.8251  & & 0.4803  & 0.4920      & 0.5351
\\
$\infty$ & & 0.5000  & 860.1797   & 3730.5741  & & 0.5000  & 0.5117      & 0.5548
\\
\noalign{\smallskip}\hline
\end{tabular}}
\\
We use symbol $I_{1s}$ to denote the absolute binding energy. The scaled binding energy $\bar{I}_{1s}$ is defined as $\bar{I}_{1s}=I_{1s}/Z^{2}$ where $Z$ is the atomic number. It is seen that $\bar{I}_{1s}$ is identical to $I_{1s}$ for H atom since $Z=1$.
\end{table}

\subsection{Photoionization cross section\label{sec3A}}

\begin{table}
\caption{Total photoionization cross sections in megabarn (Mb) for the ground state of hydrogen atom in the electric-dipole approximation with scaled shielding lengths $\Delta=5$ and $\Delta=20$. }
\label{table cross section}
\begin{tabular}{lccccc}
\hline\noalign{\smallskip}
$\Delta$ & $\omega$ & Present & \cite{Zhao} & \cite{Xie}
\\
\noalign{\smallskip}\hline\noalign{\smallskip}
5  & 0.453 & 7.2723[$+0$]  & 7.2724[$+0$]  & 7.2724[$+0$]
\\
   & 0.455 & 7.1904[$+0$]  & 7.1904[$+0$]  & 7.1904[$+0$]
\\
   & 1     & 8.6060[$-1$]  & 8.6058[$-1$]  & 8.6060[$-1$]
\\
   & 10    & 7.7809[$-4$]  & 7.7848[$-4$]  & 7.7742[$-4$]
\\
\noalign{\smallskip}\hline\noalign{\smallskip}
20 & 0.453 & 8.2359[$+0$]  & 8.2328[$+0$] & 8.2404[$+0$]
\\
   & 0.455 & 8.0247[$+0$]  & 8.0223[$+0$] & 8.0221[$+0$]
\\
   & 1     & 9.2627[$-1$]  & 9.2627[$-1$] & 9.2640[$-1$]
\\
   & 10    & 8.1699[$-4$]  & 8.1693[$-4$] & 8.1632[$-4$]
\\
\noalign{\smallskip}\hline
\end{tabular}
\\
Numbers in the brackets denote powers of 10.
\end{table}

Table \ref{table cross section} shows our results for the total photoionization cross sections of the H atom in its $^{2}S_{1/2}$ ground-state within $E1$ approximation, where comparisons with calculations of \cite{Zhao} and \cite{Xie} are made. It is observed that the agreements among different calculations are good with discrepancies less than $0.1\%$. The origin of the slight discrepancies are probable due to different numerical schemes adopted in distinct approaches. It is found that, as expected, summarized multipole contributions other than $E1$ contributions are smaller than $0.01\%$ of the exact cross sections for neutral H atom in the photon energy range of interest, from $\omega=1.0I_{1s}$ to $2.0I_{1s}$. Nevertheless, it is crucial to take multipole effects beyond $E1$ approximation into account for Nb$^{40+}$ and Pb$^{81+}$ because of their relatively much higher binding energies which require high incident photon energies to induce photoelectrons.

\begin{table*}
\caption{Near-threshold total photoionization cross sections $\sigma_{1s}$ in Mb for H-like ions Nb$^{40+}$ and Pb$^{81+}$ in the ground states with scaled shielding lengths $\Delta=2$, $10$, $50$ and $\infty$. Numbers in brackets denote powers of 10. }
\label{table summary0}
\setlength{\tabcolsep}{5.5mm}{  
\begin{tabular}{lccccccccc}
\hline\noalign{\smallskip}
 & & \multicolumn{3}{c}{ Nb$^{40+}$} & & \multicolumn{3}{c}{Pb$^{81+}$}
\\
\noalign{\smallskip}\hline\noalign{\smallskip}
$\Delta$ & & $\sigma_{1s}^{\textrm{(Exact)}}$ & $\sigma_{1s}^{(E1)}$ & $R$ & & $\sigma_{1s}^{\textrm{(Exact)}}$ & $\sigma_{1s}^{(E1)}$ & $R$
\\
\noalign{\smallskip}\hline\noalign{\smallskip}
2        & & 8.5300[$-3$]  & 8.4989[$-3$]  & 0.36\%  & & 1.6823[$-3$]  & 1.6526[$-3$]  & 1.76\%
\\
10       & & 4.2611[$-3$]  & 4.2083[$-3$]  & 1.24\%  & & 8.5206[$-4$]  & 8.0969[$-4$]  & 4.97\%
\\
50       & & 3.6654[$-3$]  & 3.6058[$-3$]  & 1.63\%  & & 7.3554[$-4$]  & 6.8949[$-4$]  & 6.26\%
\\
$\infty$ & & 3.5257[$-3$]  & 3.4639[$-3$]  & 1.75\%  & & 7.0837[$-4$]  & 6.6121[$-4$]  & 6.66\%
\\
\noalign{\smallskip}\hline
\end{tabular}}
\\
Deviation $R\equiv 100\% \times[\sigma_{1s}^{\textrm{(Exact)}}-\sigma_{1s}^{(E1)}]/\sigma_{1s}^{\textrm{(Exact)}}$, where $\sigma_{1s}^{(E1)}$ is the total cross section obtained within the $E1$ approximation while $\sigma_{1s}^{\textrm{(Exact)}}$ is the fully converged total cross section achieved with higher multipole contributions included. The photoelectron energy is assigned as barely as 0.5 a.u. to reflect circumstances of photoionization processes virtually happening at the thresholds.
\end{table*}

To explicate the importance of multipole effects on near-threshold photoionization processes in ions Nb$^{40+}$ and Pb$^{81+}$ in the ground-states, we pick a photoelectron energy as low as 0.5 a.u. for obtaining total photoionization cross sections $\sigma_{1s}$ to clarify this point. In Table \ref{table summary0}, we present the achieved ¡°exact¡± $\sigma_{1s}^{\textrm{(Exact)}}$ by summing over all multipoles with notable contributions together with the $\sigma_{1s}^{(E1)}$ within the $E1$ approximation for H-like ions Nb$^{40+}$ and Pb$^{81+}$. In addition, a deviation $R$ which stands for the measure of relative discrepancy between $\sigma_{1s}^{(E1)}$ and $\sigma_{1s}^{\textrm{(Exact)}}$ is given as well. In precise notation, the deviation $R$ is define as $R\equiv[\sigma_{1s}^{\textrm{(Exact)}}-\sigma_{1s}^{(E1)}]/\sigma_{1s}^{\textrm{(Exact)}}$. The scaled shielding length are chosen at $\Delta = ZD =2$, $10$, $50$ and $\infty$. The results in Table \ref{table summary0}, with $R$ ranging from 0.36\% to 6.66\%, clearly demonstrate that multipole effects beyond the E1 approximation actually affect significantly on the photoionization processes even occurring virtually at the ionization threshold. It is worth noticed that, in practice, we use (\ref{reduced cross section}) to achieve converged total cross sections by summing over multipoles $(E1,M1)\rightarrow(E5,M5)$ and $(E1,M1)\rightarrow(E10,M10)$ for Nb$^{40+}$ and Pb$^{81+}$, correspondingly. The contributions from all left over higher multipoles are estimated to be less than 0.01\% of the converged results.

\begin{table*}
\caption{Total photoionization cross sections $\sigma_{1s}$ in Mb for H-like ions Nb$^{40+}$ and Pb$^{81+}$ in the ground states with scaled shielding lengths $\Delta=2$, $10$, $50$ and $\infty$. }
\label{table summary} 
\setlength{\tabcolsep}{4.5mm}{ 
\begin{tabular}{lcccccccccc}
\hline\noalign{\smallskip}
 & & & \multicolumn{3}{c}{ Nb$^{40+}$} & & \multicolumn{3}{c}{Pb$^{81+}$}
\\
\noalign{\smallskip}\hline\noalign{\smallskip}
$\Delta$ & $\bar{\omega}$ & & $\sigma_{1s}^{\textrm{(Exact)}}$ & $\sigma_{1s}^{(E1)}$ & $R$ & & $\sigma_{1s}^{\textrm{(Exact)}}$ & $\sigma_{1s}^{(E1)}$ & $R$
\\
\hline
2        & 1.01  & & 8.4136[$-3$]  & 8.3825[$-3$]  & 0.37\%  & & 1.6534[$-3$]  & 1.6237[$-3$]  & 1.80\%
\\
         & 1.30  & & 5.4120[$-3$]  & 5.3799[$-3$]  & 0.59\%  & & 1.0359[$-3$]  & 1.0053[$-3$]  & 2.95\%
\\
         & 1.60  & & 3.7054[$-3$]  & 3.6735[$-3$]  & 0.86\%  & & 6.9423[$-4$]  & 6.6435[$-4$]  & 4.30\%
\\
         & 2.00  & & 2.4248[$-3$]  & 2.3942[$-3$]  & 1.26\%  & & 4.4445[$-4$]  & 4.1646[$-4$]  & 6.30\%
\\
\noalign{\smallskip}\hline\noalign{\smallskip}
10       & 1.01  & & 4.1676[$-3$]  & 4.1143[$-3$]  & 1.28\%  & & 8.3225[$-4$]  & 7.8934[$-4$]  & 5.16\%
\\
         & 1.30  & & 2.2467[$-3$]  & 2.1876[$-3$]  & 2.63\%  & & 4.4944[$-4$]  & 4.0142[$-4$]  & 10.69\%
\\
         & 1.60  & & 1.3280[$-3$]  & 1.2735[$-3$]  & 4.11\%  & & 2.6753[$-4$]  & 2.2359[$-4$]  & 16.42\%
\\
         & 2.00  & & 7.4214[$-4$]  & 6.9661[$-4$]  & 6.13\%  & & 1.5196[$-4$]  & 1.1588[$-4$]  & 23.74\%
\\
\noalign{\smallskip}\hline\noalign{\smallskip}
50       & 1.01  & & 3.5777[$-3$]  & 3.5173[$-3$]  & 1.69\%  & & 7.1752[$-4$]  & 6.7079[$-4$]  & 6.51\%
\\
         & 1.30  & & 1.8405[$-3$]  & 1.7748[$-3$]  & 3.57\%  & & 3.7603[$-4$]  & 3.2408[$-4$]  & 13.82\%
\\
         & 1.60  & & 1.0499[$-3$]  & 9.9198[$-4$]  & 5.52\%  & & 2.1907[$-4$]  & 1.7315[$-4$]  & 20.96\%
\\
         & 2.00  & & 5.6678[$-4$]  & 5.2088[$-4$]  & 8.10\%  & & 1.2212[$-4$]  & 8.5955[$-5$]  & 29.62\%
\\
\noalign{\smallskip}\hline\noalign{\smallskip}
$\infty$ & 1.01  & & 3.4393[$-3$]  & 3.3768[$-3$]  & 1.82\%  & & 6.9079[$-4$]  & 6.4291[$-4$]  & 6.93\%
\\
         & 1.30  & & 1.7464[$-3$]  & 1.6790[$-3$]  & 3.86\%  & & 3.5926[$-4$]  & 3.0628[$-4$]  & 14.75\%
\\
         & 1.60  & & 9.8693[$-4$]  & 9.2830[$-4$]  & 5.94\%  & & 2.0820[$-4$]  & 1.6185[$-4$]  & 22.26\%
\\
         & 2.00  & & 5.2810[$-4$]  & 4.8235[$-4$]  & 8.66\%  & & 1.1556[$-4$]  & 7.9468[$-5$]  & 31.23\%
\\
\noalign{\smallskip}\hline
\end{tabular}}
\\
The notations $\sigma_{1s}^{(E1)}$ and $\sigma_{1s}^{\textrm{(Exact)}}$ and $R$ are the same as those defined in Table \ref{table summary0}. The photoelectron energy is assigned as barely as 0.5 a.u. to reflect circumstances of photoionization processes virtually happening at the thresholds. The reduced photon energy $\bar{\omega}\equiv\omega/I_{1s}$ with $\omega$ and $I_{1s}$ being the true photon and binding energies, respectively, is as introduced in the context.
\end{table*}

To examine the multipole effects on the total photoionization cross sections of ions Nb$^{40+}$ and Pb$^{81+}$ at incident photon energies departing away from ionization threshold, here we introduce the reduced photon energy $\bar{\omega}$ by the definition $\bar{\omega}\equiv\omega/I_{1s}$, with $\omega$ being the true photon energy and $I_{1s}$ the binding energy. It is emphasized that $\bar{\omega}$ is dimensionless and we multiply binding energy $I_{1s}$ by $\bar{\omega}$ to give the true photon energy $\omega$. In other words, $\bar{\omega}$ corresponds to $\omega$ in unit of binding energy $I_{1s}$. In the present study, the reduced photon energies of interest are in the region between 1.01 and 2.00 corresponding to true photon energies ranging from $1.01I_{1s}$ and $2.00I_{1s}$. With the same set of scaled shielding lengths for Table \ref{table summary0}, in Table \ref{table summary}, we present $\sigma_{1s}^{\textrm{(Exact)}}$, $\sigma_{1s}^{(E1)}$ and $R$ for ions Nb$^{40+}$ and Pb$^{81+}$ at exemplary $\bar{\omega}=1.01$, $1.3$, $1.6$ and $2.0$. As we can see, multipole contributions to total photoionization cross sections are raised with increasing $\bar{\omega}$.

\begin{figure*}
\includegraphics[width=0.49\textwidth]{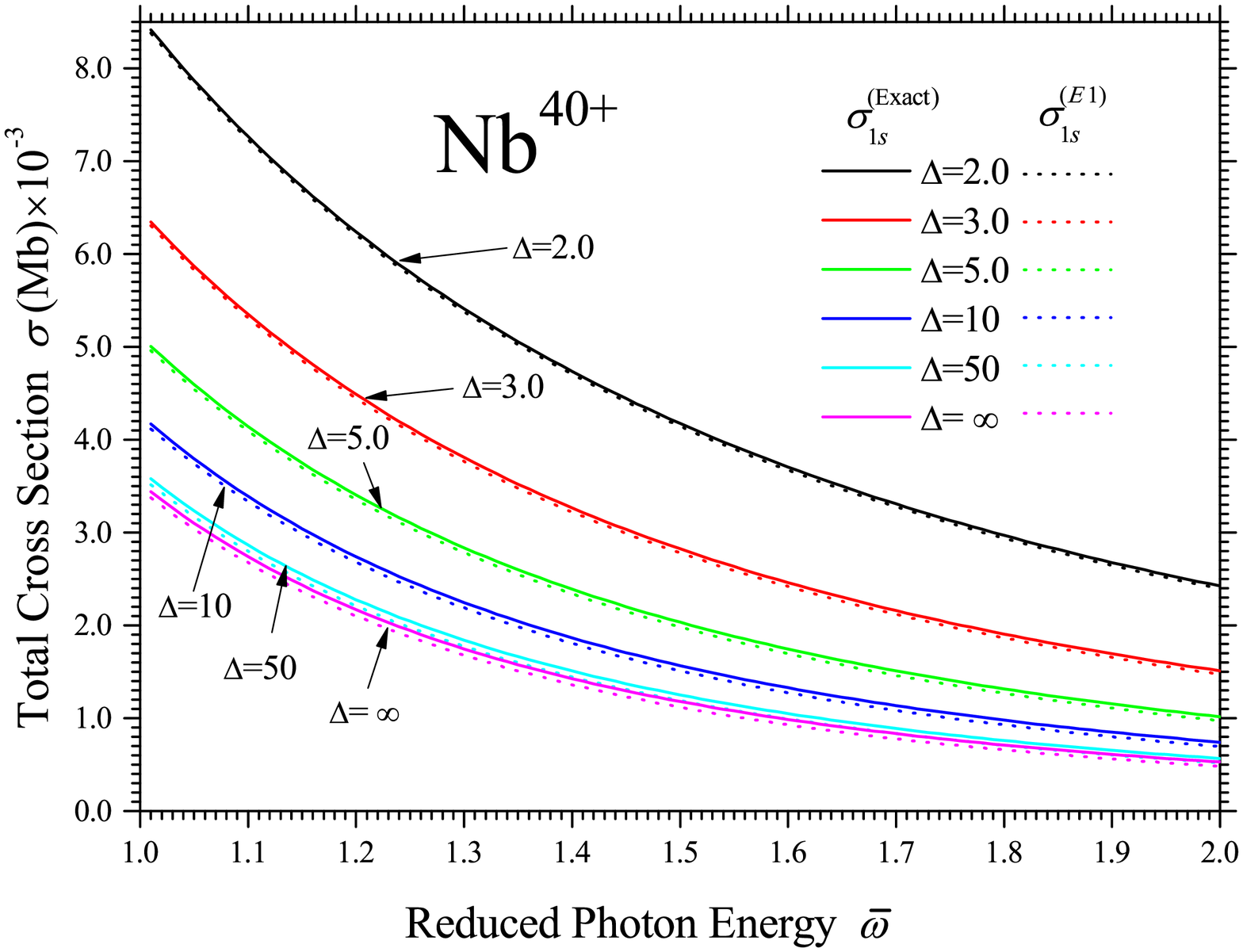}
\includegraphics[width=0.49\textwidth]{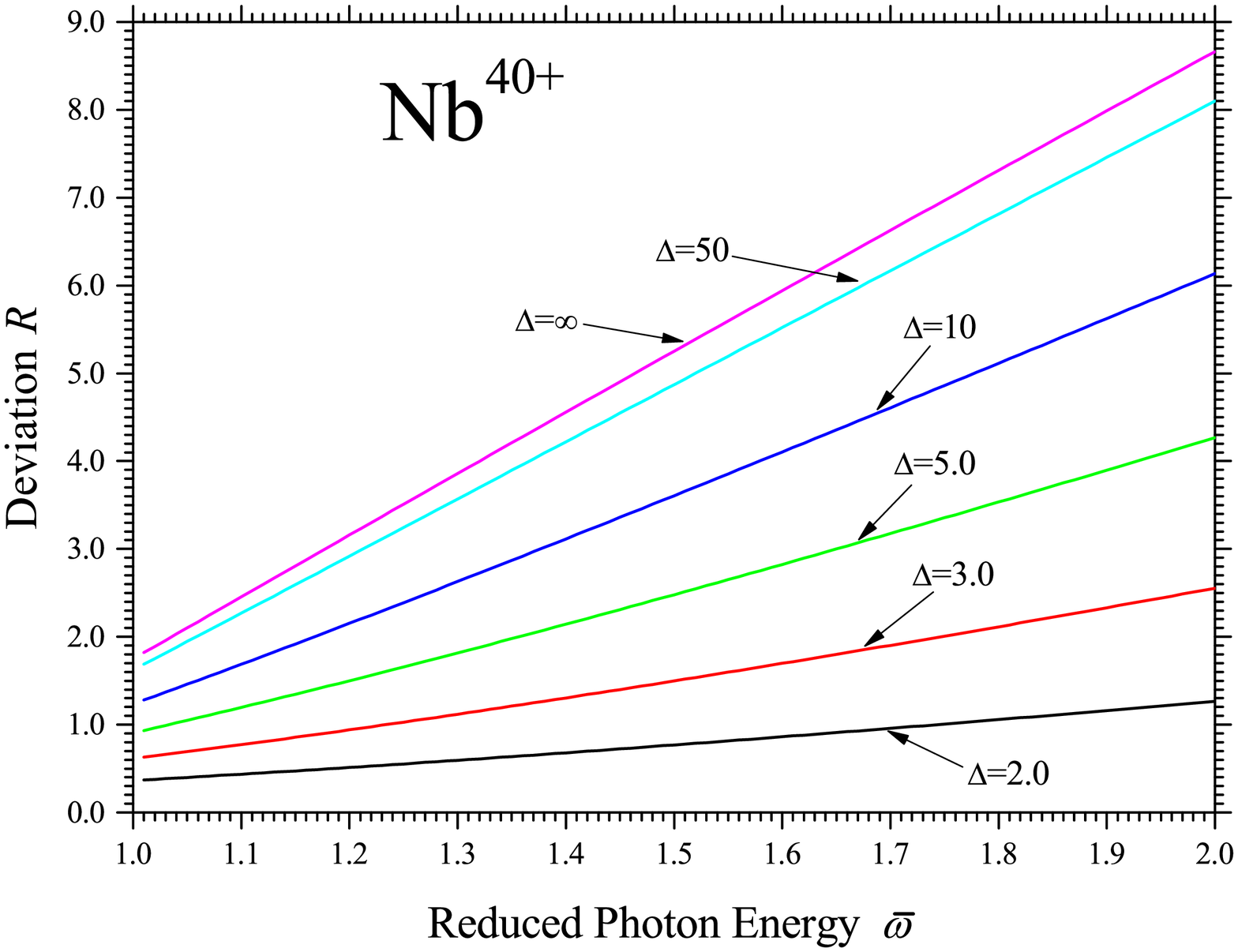}
\caption{\label{multipole_contribution41} Left panel: plot of total cross sections of the H-like ion Nb$^{40+}$ in Mb with various Debye lengths $D$ indicated in the plot. Solid lines are exact cross sections $\sigma_{1s}^{\textrm{(Exact)}}$ while dot lines correspond to results within the $E1$ approximation, denoted by $\sigma_{1s}^{(E1)}$. Right panel: plot of deviations $R$. The deviation $R$ is a measure of relative discrepancy between $\sigma_{1s}^{\textrm{(Exact)}}$ and $\sigma_{1s}^{(E1)}$. In precise notation, $R\equiv[\sigma_{1s}^{\textrm{(Exact)}}-\sigma_{1s}^{(E1)}]/\sigma_{1s}^{\textrm{(Exact)}}$. It is remarked that $\sigma_{1s}^{\textrm{(Exact)}}$, with uncertainty smaller than $0.01\%$, is achieved by summing over contributions from transitions induced through electric multipoles $E1$ to $E5$ and magnetic multipoles $M1$ to $M5$.}
\end{figure*}

\begin{figure*}
\centering
\includegraphics[width=0.49\textwidth]{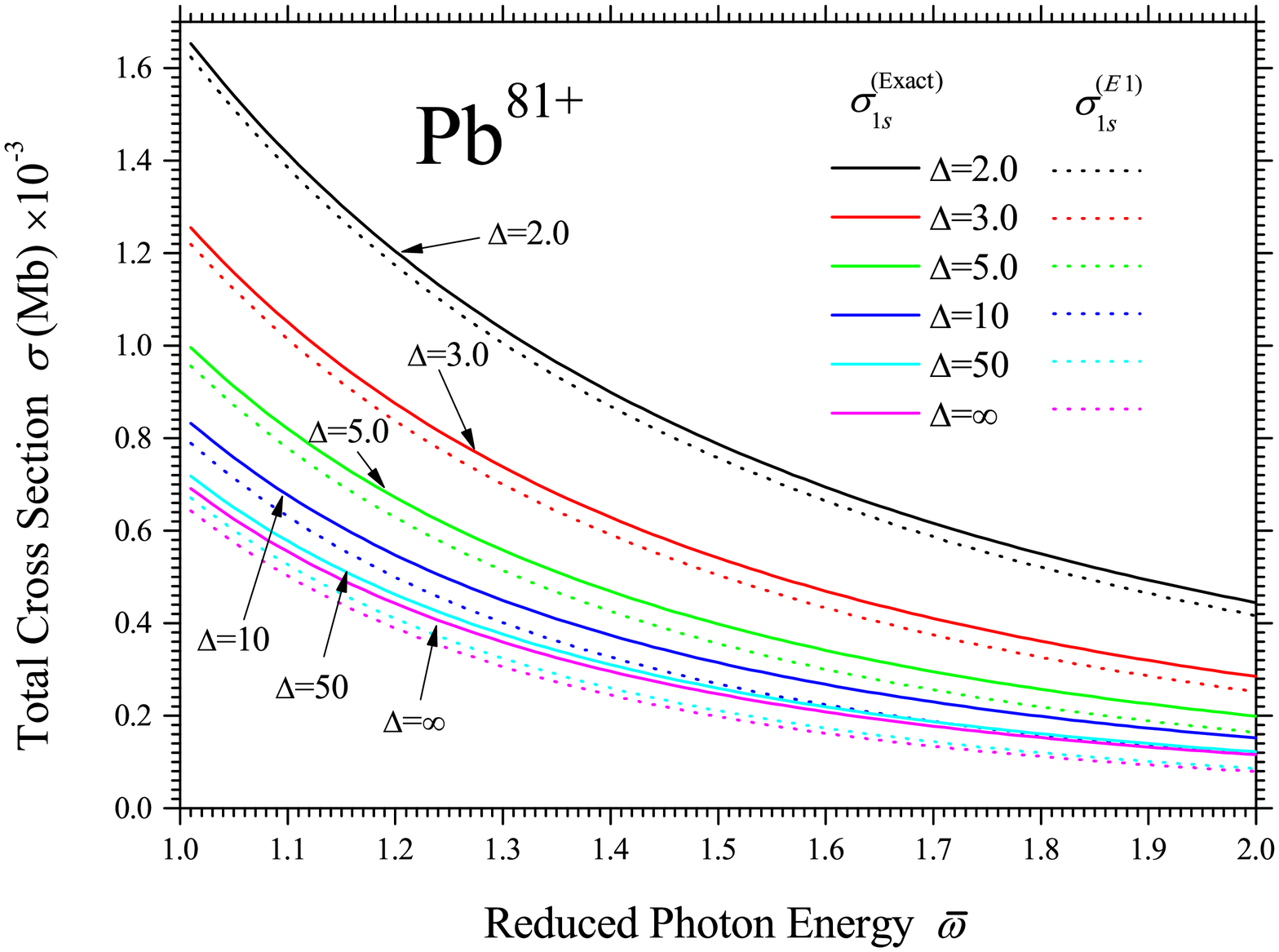}
\includegraphics[width=0.49\textwidth]{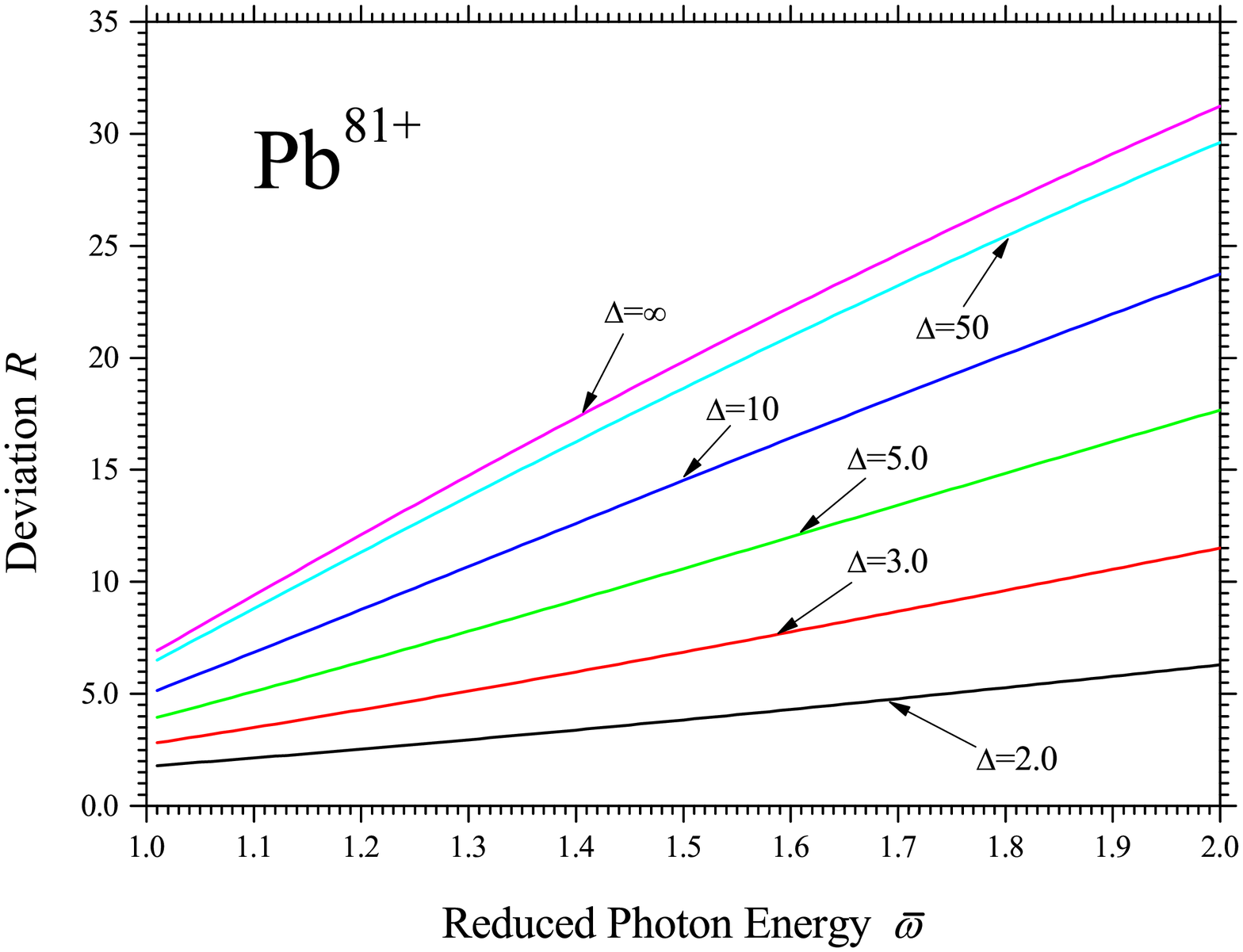}
\caption{\label{multipole_contribution82} Left panel: plot of total cross sections of the H-like ion Pb$^{81+}$ in Mb with various Debye lengths $D$ indicated in the plot. Solid lines are exact cross sections $\sigma_{1s}^{\textrm{(Exact)}}$ while dot lines correspond to results within the $E1$ approximation, denoted by $\sigma_{1s}^{(E1)}$. Right panel: plot of deviations $R$. The deviation $R$ is a measure of relative discrepancy between $\sigma_{1s}^{\textrm{(Exact)}}$ and $\sigma_{1s}^{(E1)}$. It is remarked that $\sigma_{1s}^{\textrm{(Exact)}}$, with uncertainty smaller than $0.01\%$, is achieved by summing over contributions from transitions induced through electric multipoles $E1$ to $E10$ and magnetic multipoles $M1$ to $M10$.}
\end{figure*}

In Fig. \ref{multipole_contribution41}, we plot the total photoionization cross sections against the reduced photon energy for Nb$^{40+}$. Besides, a similar plot for total photoionization cross sections of Pb$^{81+}$ is presented in Fig. \ref{multipole_contribution82}. The appearing resemblance between Fig. \ref{multipole_contribution41} and Fig. \ref{multipole_contribution82} is owing to the advantage of employing the reduced photon energy as an alternative to the true photon energy; therefore, possible scaling between the results of Nb$^{40+}$ and Pb$^{81+}$ is implied. In Fig. \ref{multipole_contribution41} and Fig. \ref{multipole_contribution82}, respectively, the left panels show the exact cross sections $\sigma_{1s}^{\textrm{(Exact)}}$ together with the $E1$ approximated cross sections $\sigma_{1s}^{(E1)}$, in the meantime the deviations $R$ are depicted in the right panels. From the right panels in Fig. \ref{multipole_contribution41} and Fig. \ref{multipole_contribution82} individually, it is evident that non-electric dipole contributions are enhanced with shielding lengths $\Delta$ prolonged, as we may observe from the consonantly enlarged deviations $R$. Furthermore, it is observed that the deviations $R$ depend approximately linearly on the reduced photon energies for ion Nb$^{40+}$. On the other hand,  an approximately linear dependence of $R$ on $\bar{\omega}$ is seen for ion Pb$^{81+}$ as well.

Attentions are also paid to the influences of shielding lengths $\Delta$ on the linearity property of $R$ as functions of $\bar{\omega}$, we find the following two aspects in consequence: (1) in strong shielding case with small $\Delta$, the better the linearity relations. (2) The greater the shielding lengths, the linearity is mildly distorted and qualitatively correct. Although Fig. \ref{multipole_contribution41} and Fig. \ref{multipole_contribution82} resemble each other, we discover that $\sigma_{1s}^{(E1)}[\textrm{Nb}^{40+}]/\sigma_{1s}^{(E1)}[\textrm{Pb}^{81+}]$ is on average 5.02 and 5.00, respectively, close to the threshold and at the $\bar{\omega}=2.0$ end. Moreover, larger $R$ is unfolded for Pb$^{81+}$ in comparison with Nb$^{40+}$, which embodies the fact that multipole effects should be included for Pb$^{81+}$ are in five orders higher than those should be included for Nb$^{40+}$, as a result of higher binding energy combined with more prominent relativistic effects in Pb$^{81+}$. We also inspect the ratio $R[\textrm{Pb}^{81+}]/R[\textrm{Nb}^{40+}]$ for various $\Delta$ in the vicinity of the ionization threshold and at the $\bar{\omega}=2.0$ end. It is found that the ratios are between 3.81 and 4.86 with an average of 4.21 surrounding the ionization threshold whereas they range from 3.61 to 5.00 with a mean of 3.90 at the $\bar{\omega}=2.0$ end, an interesting outcome raised to be compared with the ratio $Z^{2}[\textrm{Pb}]/Z^{2}[\textrm{Nb}]=82^{2}/41^{2}=4.00$. To reveal how plasma shielding affect multipole effects, the slopes of $R$, symbolized by $m$, for each $\Delta$ are also best estimated. With $\Delta$ varying from 2 to $\infty$, the slope $m$ monotonically increases from 0.91 to 6.92 for Nb$^{40+}$, and it rises from 4.57 to 24.6 for Pb$^{81+}$in parallel. Since the slope $m$ and $\Delta$ are positively correlated, it is illustrated that multipole effects are softened as plasmas shielding effects being intensified. Another interesting indicator for probing is the ratio $m[\textrm{Pb}^{81+}]/m[\textrm{Nb}^{40+}]$ for distinct $\Delta$. The ratios $m[\textrm{Pb}^{81+}]/m[\textrm{Nb}^{40+}]$ scope from 3.56 to 5.04 with an average of 4.12, an aftermath comparable to that of $R[\textrm{Pb}^{81+}]/R[\textrm{Nb}^{40+}]$.

Aside from the aforementioned features, detailed examinations of present calculations with $\Delta=1$ show that the contributive portions arising from the non-electric dipole effects to the total ionization cross section are indeed smaller than $1\%$ of the exact result for both ions Nb$^{40+}$ and Pb$^{81+}$. It means that $E1$ approximation is satisfactory for acquiring a total photoionization cross section accurate to $99\%$ under the $\Delta=1$ conditions corresponding to ultra plasma shielding. In ultra plasma shielding cases with $\Delta$ being close to 1, it is plausible to anticipate that, for incident photon energies of interest in the present study, frameworks within $E1$ approximation will be considerably appropriate since the nuclear charge is significantly shielded off from the plasma environments.

\subsection{Angular distribution and spin polarization of photoelectrons within the $E1$ approximation\label{sec3B}}

In this subsection, within the $E1$ approximation, we investigate the effects of plasma shielding on the angular distribution and total spin polarization parameters $\beta_{1s}$ and $\delta_{1s}$ of photoelectrons using various scaled shielding lengths $\Delta$ for H atom and H-like ions Nb$^{40+}$ and Pb$^{81+}$. Similar to the total photoionization cross section $\sigma_{1s}$, in situations of high incident photon energy, it is worthwhile to point out that high-multipole transitions beyond the $E1$ approximation must also be considered. This is due to the fact that the dominant $E1$ amplitudes will interfere coherently with amplitudes arising from high-multipole transitions to give angular distribution and total spin polarization parameters. While $\sigma_{1s}$ is given by summing incoherently over squares of distinct multipole transition amplitudes, in contrast asymmetry parameter $\beta_{1s}$ and spin polarization parameter $\delta_{1s}$ are obtained from summations over squares of terms involving interference among different multipole amplitudes. Because of the persistent interference terms, expressions of $\beta_{1s}$ and $\delta_{1s}$ are more complicate compared to $\sigma_{1s}$. In practice, the higher the multipole transition amplitudes to be included, the more the complexities in the expressions of $\beta_{1s}$ and $\delta_{1s}$. Here we restrict ourselves to the $E1$ approximation, further attempts to include the multipole interference effects are under our development.

As it is discussed in Sec. \ref{sec2A}, the angular distribution and total spin polarization parameters are 2.0 and 0.0, respectively, in the non-relativistic limit. Therefore, the parameters $\beta_{1s}$ and $\delta_{1s}$ for H atom are not given because H atom manifest itself in a rather non-relativistic behavior. Deviations of $\beta_{1s}$ and $\delta_{1s}$ away from 2.0 and 0.0 expose the onset of relativistic effects leading to spin-orbit splitting of the amplitudes $D_{1/2}$ and $D_{3/2}$ introduced in (\ref{sigma expression}) to (\ref{delta expression}). Due to the splitting of the amplitudes $D_{1/2}$ and $D_{3/2}$ activated by the spin-orbit couplings, $\beta_{1s}$ and $\delta_{1s}$ depart from their non-relativistic limits as a result. Since $D_{1/2}$ and $D_{3/2}$ are energy dependent, $\beta_{1s}$ and $\delta_{1s}$ depend on energy as well. As for the photoionization cross section, we explore the tendencies of $\beta_{1s}$ and $\delta_{1s}$ with respect to the reduced photon energy in the following.

\begin{figure}
\includegraphics[width=0.55\textwidth]{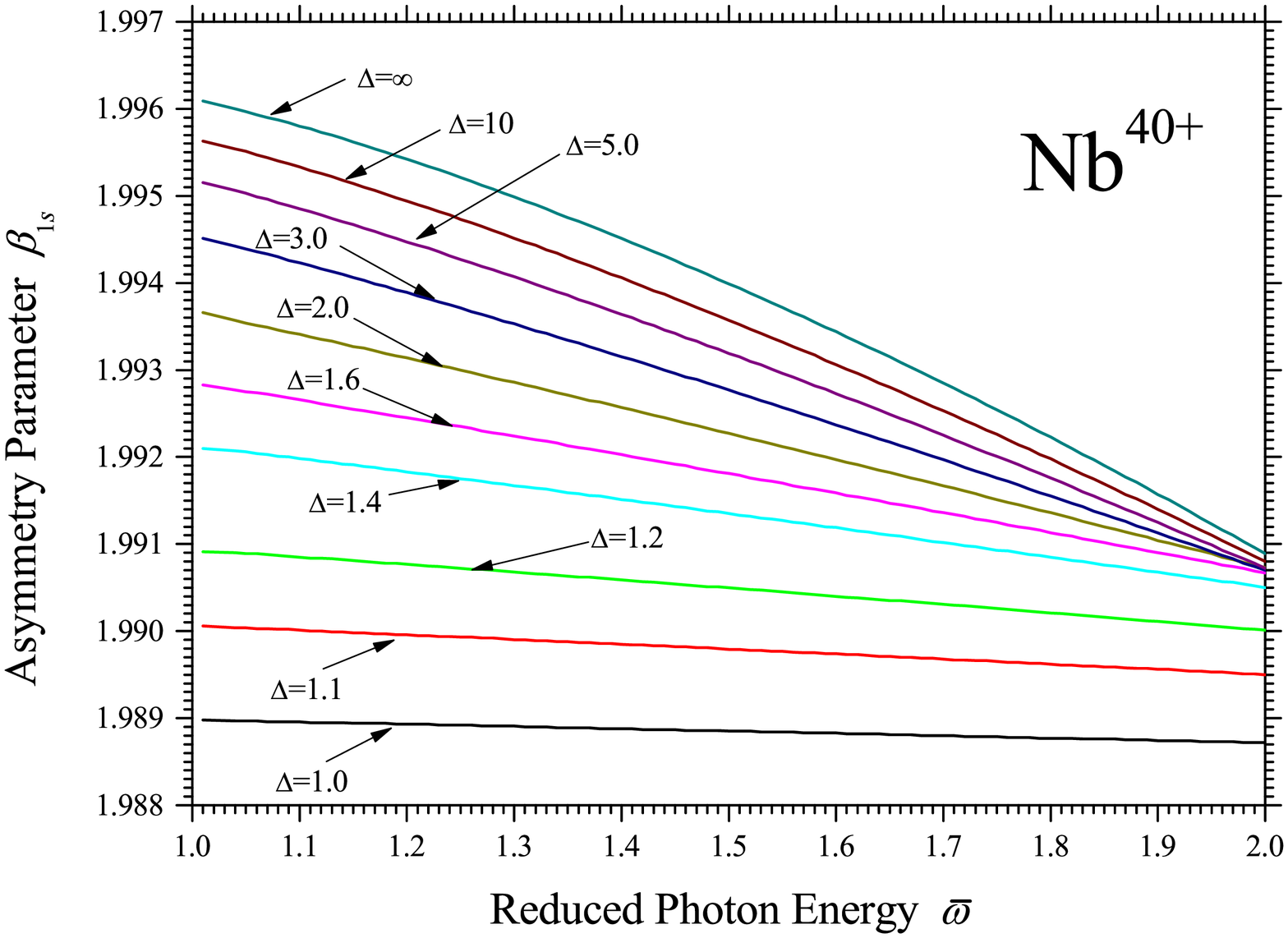}
\caption{\label{beta1} Angular asymmetry parameter $\beta_{1s}$ versus reduced photon energy $\bar{\omega}$ with different scaled shielding lengths $\Delta$ in the hydrogen-like Nb$^{40+}$ ions.}
\includegraphics[width=0.55\textwidth]{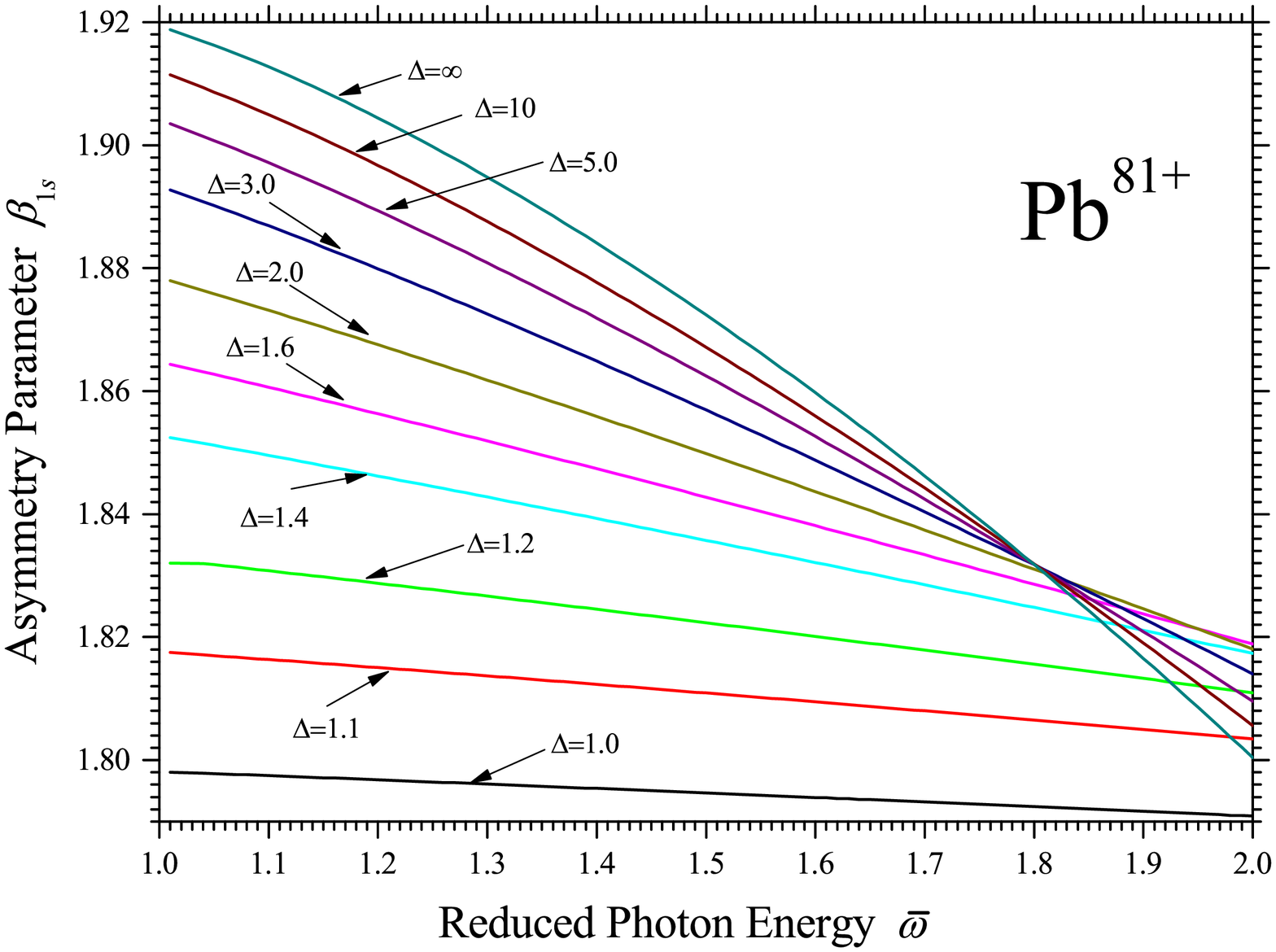}
\caption{\label{beta2} Angular asymmetry parameter $\beta_{1s}$ versus reduced photon energy $\bar{\omega}$ with different scaled shielding lengths $\Delta$ in the hydrogen-like Pb$^{81+}$ ions.}
\end{figure}

In Fig. \ref{beta1} and Fig. \ref{beta2}, we plot $\beta_{1s}$ as functions of the reduced photon energy $\bar{\omega}$ for Nb$^{40+}$ and Pb$^{81+}$, respectively. It is seen that $\beta_{1s}$ apparently diverge from 2.0 for Pb$^{81+}$ owing to pronounced relativistic effects. In comparison, $\beta_{1s}$ for Nb$^{40+}$ deviates very slightly from 2.0 because relativistic effects are not as noticeable as in the case of Pb$^{81+}$. We see that $\beta_{1s}$ decrease monotonically against $\bar{\omega}$. A general trend to be observed is that $\beta_{1s}$ bends farther away from 2.0 for higher reduced photon energy equivalent to more energetic photoelectrons, a manifestation agrees with the common understanding of relativity. Furthermore, the effects of plasma shielding on $\beta_{1s}$ are demonstrated in Fig. \ref{beta1} and Fig. \ref{beta2} where the dependence on shielding lengths $\Delta$ are clearly shown. We first witness that plasma shielding seems to boost the influence of spin-orbit couplings on the asymmetry parameter. As it is displayed, $\beta_{1s}$ separates more remote from 2.0 in the course of diminishing the shielding length at a specific $\bar{\omega}$. We also find that strong shielding will cause the descending rate $d\beta_{1s}/d\bar{\omega}$ to be flattened. In ultra-shielding case with $\Delta\approx1.0$, $\beta_{1s}$ is inclined to be almost a constant with respect to $\bar{\omega}$ in the entire energy region of interest. An additional feature worth marking is that $\beta_{1s}$ with $\Delta\geq2.0$ tend toward to coincide at a particular $\bar{\omega}\approx1.8$ for Pb$^{81+}$. Similar coincidence is supposed to surface at a higher $\bar{\omega}>2.0$ outside the purview of the present calculations for Nb$^{40+}$. The showing up of the coincidence at a special energy point originates form the influences interplayed intricately by the relativistic effects and plasma shielding.

\begin{figure}
\includegraphics[width=0.55\textwidth]{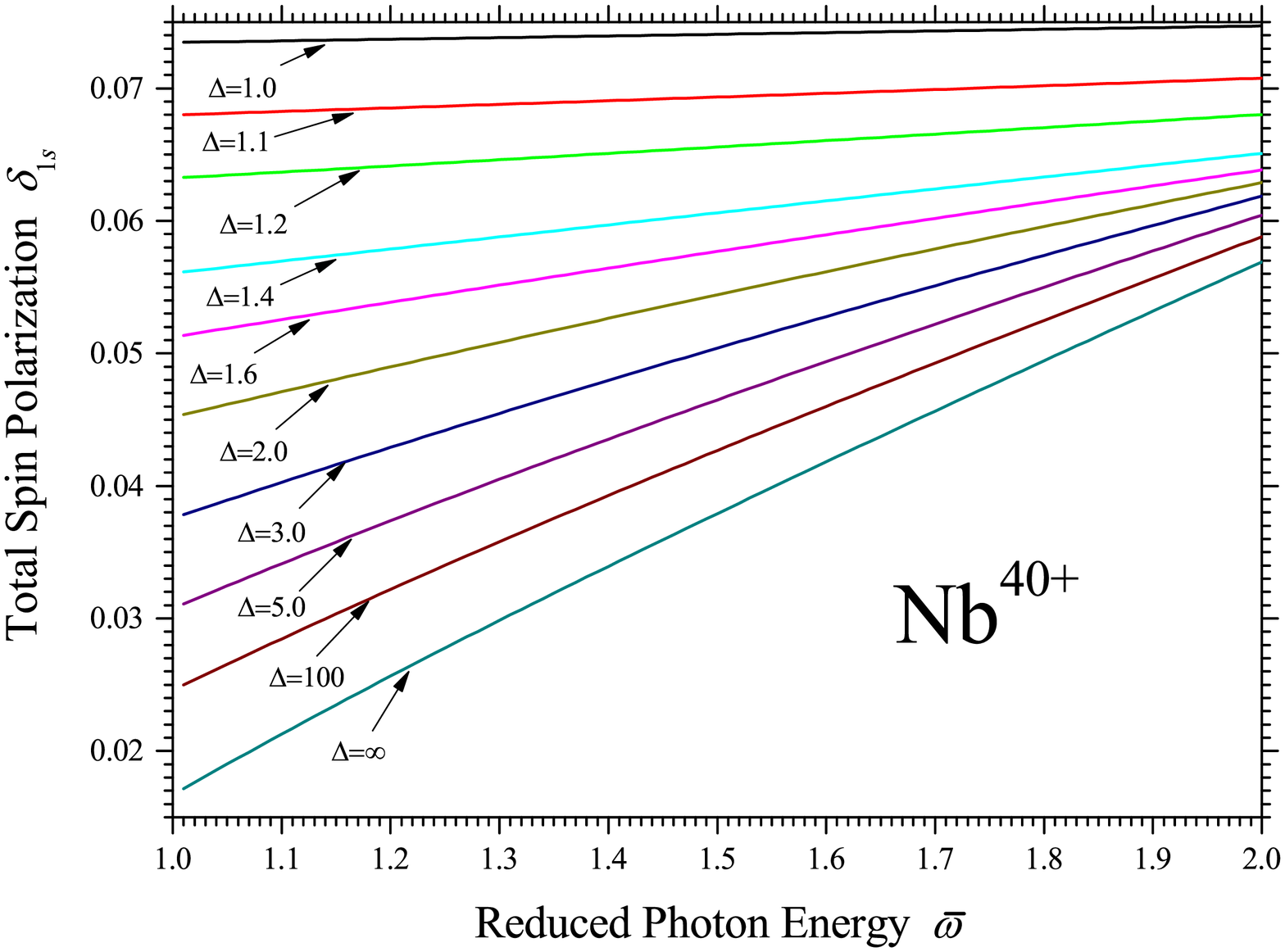}
\caption{\label{delta1} Total spin polarization parameter $\delta_{1s}$ versus reduced photon energy $\bar{\omega}$ with different scaled shielding lengths $\Delta$ in the hydrogen-like Nb$^{40+}$ ions.}
\includegraphics[width=0.55\textwidth]{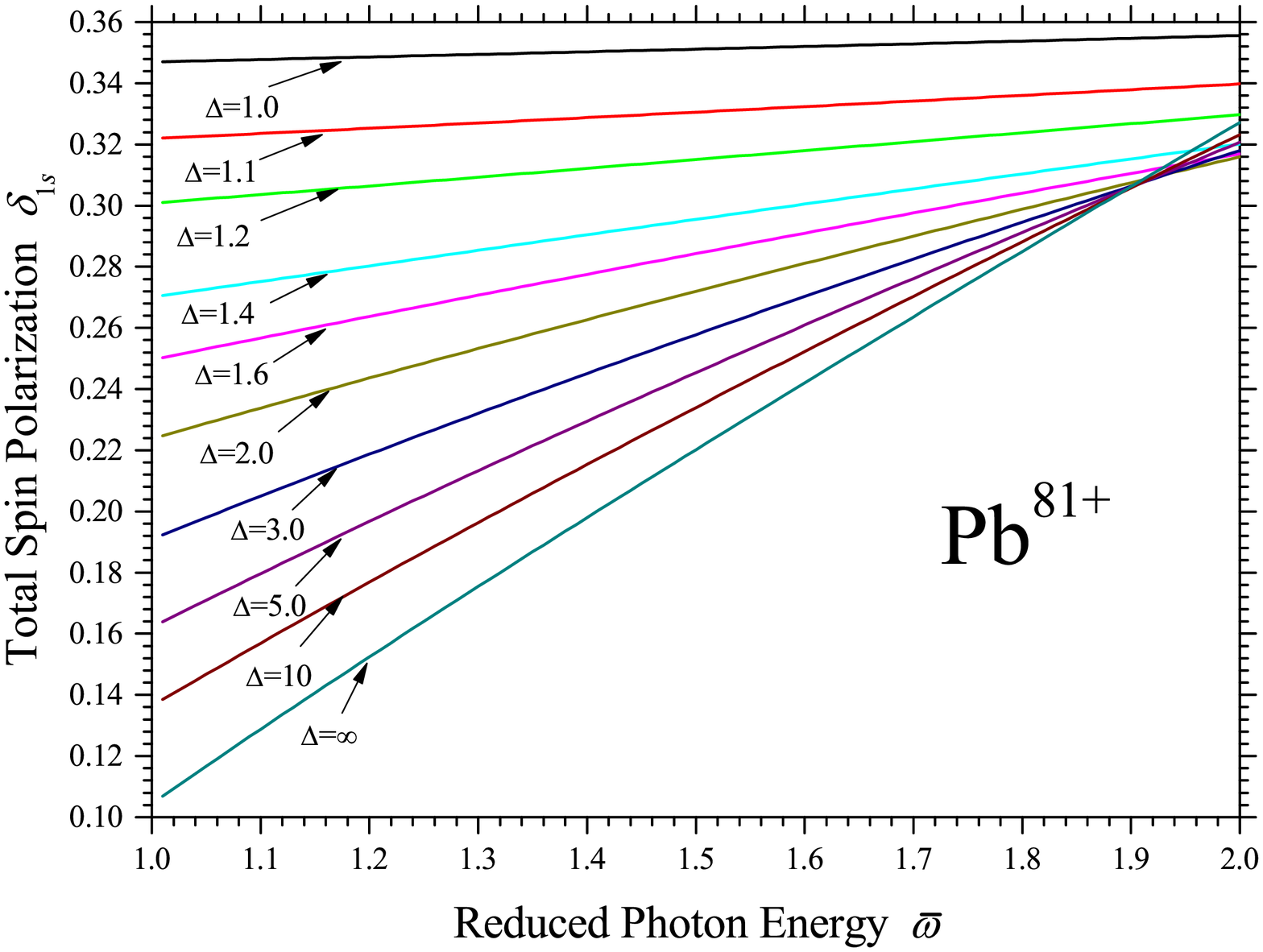}
\caption{\label{delta2} Total spin polarization parameter $\delta_{1s}$ versus reduced photon energy $\bar{\omega}$ with different scaled shielding lengths $\Delta$ in the hydrogen-like Pb$^{81+}$ ions.}
\end{figure}

The spin polarization of the total photoelectron flux is given by $P_{tot}=\pm\delta$, where the $\pm$ signs refer to incident photons with helicity $\pm1$ or the right (+) and left ($-$) circular polarization. The spin-polarization parameters $\delta_{1s}$ provides the important information about the transfer of photon polarization to photoelectron polarization. Fig. \ref{delta1} and Fig. \ref{delta2}, respectively, show the total spin-polarization parameters $\delta_{1s}$ against the reduced photon energy $\bar{\omega}$ for ions Nb$^{40+}$ and Pb$^{81+}$ with diverse scaled shielding lengths. As we may observe from Fig. \ref{delta1} and Fig. \ref{delta2}, relativistic and plasma shielding effects combined together cause $\delta_{1s}$ to reveal feature patterns in accord with those exhibited by $\beta_{1s}$. First, compared to Nb$^{40+}$, Pb$^{81+}$ displays larger $\delta_{1s}$ by virtue of more dramatic relativistic effects. Second, $\delta_{1s}$ notably differs from 0.0 for highly energetic photoelectrons. Third, the ascending rate $d\delta_{1s}/d\bar{\omega}$ is lowered down by lessened $\Delta$, and eventually becomes nearly flat in the whole energy range as $\Delta\rightarrow1$. Fourth, for Pb$^{81+}$ there presents a specific energy point $\bar{\omega}\approx1.92$ where $\delta_{1s}$ with distinctive $\Delta\geq2.0$ appear to converge at, a similar point of crossing is assumed to emerge at a higher $\bar{\omega}>2.0$ for Nb$^{40+}$.

\section{Conclusions \label{sec:4}}

In the present study we have performed a systematic study of the photoionization processes of neutral hydrogen atom and H-like ions Nb$^{40+}$ and Pb$^{81+}$ embedded in Debye plasma environments. Several typical Debye shielding lengths are selected to explore the plasma shielding effects.

We carry out calculations to obtain total photoionization cross sections accurate to five significant figures from summing over all multipoles which contribute notably. For the H atom, it is shown that $E1$ approximation is practically appropriate; besides, the present predictions agree well with available theoretical results. For high-Z H-like ions, like Nb$^{40+}$ and Pb$^{81+}$, multipole contributions in addition to the $E1$ contribution must be included even in the case of near-threshold photoionization processes. Our analyses show that multipole effects along with relativistic effects and the plasma shielding effects are essential to provide accurate total photoionization cross sections as functions of the reduced photon energy.

Although the current results of angular distribution and spin polarization parameters of photoelectrons are valid within the $E1$ approximation and will be disturbed by interferences from high-order multipole photoionization transition amplitudes, they provide prototypical demonstrations of the influences due to plasma shielding on the angular distribution and spin polarization of photoelectrons. It is evidenced that the interplay between relativistic and plasma shielding effects does effect the angular-distribution and spin-polarization parameters. Moreover, the influence of spin-orbit couplings on these parameters is reinforced as plasma shielding is strengthened. It is noteworthy that the $E1$ approximation works well in strong shielding cases with scaled shielding lengths $\Delta\approx1$; therefore, the asymmetry and polarization corresponding to such cases are practically accurate as well.

In this study, we have taken the H-atom and H-like ions Nb$^{40+}$ and Pb$^{81+}$ as representatives for low-$Z$, medium-$Z$, and high-$Z$ elements, respectively. It is anticipated that the general characteristics unraveled for photoionization parameters including total cross section, angular distribution, and spin polarization parameters in the present calculation are applicable to all H-like ions.



\begin{acknowledgements}
The author Xugen Zheng would like to thank Long Jiang for his assistance at the preliminary stage of this study. The authors acknowledge the support by the National Natural Science Foundation of China under Grant Numbers 11474209 and 11474208
\end{acknowledgements}



\end{document}